\documentclass[journal=acsnano,manuscript=article]{achemso}

\usepackage[version=3]{mhchem} % Formula subscripts using \ce{}
\usepackage{ulem} % for strikethrough text. Useful during co author review. Added by Veeresh
\usepackage{xcolor}% for color text. Useful during co author review. Added by Veeresh

\author{Anuj Bathla}
\affiliation[Indian Institute of Technology Bombay]
{Center for Research in Nanotechnology and Science,\\ Indian Institute of Technology Bombay, Mumbai 400076, India}
\alsoaffiliation[Second University]
{Department of Electrical Engineering,\\ Indian Institute of Technology Bombay, Mumbai 400076, India.}
\author{Subrat Kumar Pradhan}
\affiliation[Indian Institute of Technology Bombay]
{Department of Electrical Engineering,\\ Indian Institute of Technology Bombay, Mumbai 400076, India.}
\author{Ajit Kumar Dash}
\affiliation[Indian Institute of Science, Bengaluru]
{Department of Physics, Indian Institute of Science, Bengaluru, 560012, India}
\author{Prabhat Anand}
\affiliation[BigPharma]
{TCS Research,Tata Consultancy Services,Bengaluru, India}
\author{M. Girish Chandra}
\affiliation[Unknown University]
{TCS Research,Tata Consultancy Services,Bengaluru, India}

\author{Kenji Watanabe}
\affiliation[Unknown University]
{Research Center for Functional Materials National Institute for Materials Science Ibaraki 305-0044, Japan}
\author{Takashi Taniguchi}
\affiliation[Unknown University]
{ International Center for Materials Nanoarchitectonics National Institute for Materials Science Ibaraki 305-0044, Japan}

\author{Akshay Singh}
\affiliation[Unknown University]
{Department of Physics, Indian Institute of Science, Bengaluru, 560012, India}
\author{Veeresh Deshpande}
\affiliation[Unknown University]
{Department of Electrical Engineering,\\ Indian Institute of Technology Bombay, Mumbai 400076, India.}
\author{Kasturi Saha}
\affiliation[Unknown University]
{Department of Electrical Engineering,\\ Indian Institute of Technology Bombay, Mumbai 400076, India.}
\email{kasturis@iitb.ac.in}
\title[An \textsf{achemso} demo]
  {In-Substrate Imaging of Diamond–hBN FET Current via Wide-Field Quantum Diamond Microscopy}

\abbreviations{IR,NMR,UV}
\keywords{American Chemical Society, \LaTeX}

\begin{document}

%\begin{tocentry}

%\end{tocentry}

\begin{abstract}
We demonstrate wide-field magnetic imaging of current flow in hydrogen-terminated diamond field-effect transistors (FETs) through in-substrate nitrogen-vacancy (NV) centers. Hydrogen termination of the diamond surface induces a two-dimensional hole gas (2DHG), while an ensemble of near-surface NV centers located $ \sim 1~\mu m$ below the surface enables noninvasive magnetic imaging of current flow with micrometer-scale spatial resolution. The FETs were electrically characterized over a range of drain–source biases $V_{ds}= 0$ to $-15V$ and gate voltages,$V_{gs}= +3$ to $-9V$ followed by in situ wide-field NV magnetometry during device operation. Magnetic field maps and reconstructed current-density distributions directly visualize current injection at the source–drain contacts and transport beneath the hBN-gated channel. Magnetic field maps reveal current-density variations in the channel region owing to non-uniformities or defects in the gate dielectric. In addition, we observe a pronounced enhancement of the drain current ($\sim 600–900 \mu A$) and a shift in the apparent threshold voltage during laser illumination, reflecting photo-induced changes in channel electrostatics. By correlating gate-dependent magnetic images with simultaneous electrical measurements, we directly link spatial current distributions to FET transfer characteristics, providing new insight into buried interface transport and non-uniform gating effects in the transistor channel. As the methodology is compatible with top-gated FETs, it can be used to map channel current distributions with micrometer resolution in emerging channel materials, such as 2D materials and wide-bandgap channels, and establish wide-field NV magnetometry as a powerful platform for probing charge transport in transistors and Van der Waals dielectric heterostructures. 

\end{abstract}

\section{Introduction}
Diamond has emerged as a promising wide-bandgap semiconductor for next-generation high-power and high-frequency electronics owing to its exceptional material properties, including a large bandgap ($\sim 5.5eV \,$), high thermal conductivity, high breakdown strength, and chemical inertness\cite{Isberg2002}. Among various approaches for forming conductive channels in diamond, the hydrogen termination of the surface (100) or (111) produces a two-dimensional hole gas (2DHG) when exposed to atmospheric adsorbates or specific surface treatments\cite{Maier2000,Kawarada1996,Strobel2004_SurfaceTransferDoping}. This 2DHG exhibits high sheet conductivity, strong thermal stability, and compatibility with field-effect transistor (FET) architectures, enabling the realization of high-performance diamond metal–oxide–semiconductor FETs (MOSFETs)\cite{Geis2018_DiamondFETs,Kawarada2017_2DHG_Diamond}.\\
The performance of hydrogen-terminated diamond FETs is strongly influenced by the quality and nature of the gate dielectric. Traditional dielectric layers such as \ce{Al2O3}, \ce{SiO2}, or \ce{HfO2} often introduce interface traps, mechanical strain, or chemically degrade surface termination, adversely affecting 2DHG mobility and threshold voltage stability.\cite{Kawarada2017_2DHG_Diamond,Pham2018_DiamondMOSCAP,Liu2013_HfO2_HdiamondFET} In contrast, hexagonal boron nitride (hBN) has emerged as a promising dielectric material due to its atomic flatness, chemical inertness, large bandgap ($\sim 6 eV \,$), and absence of dangling bonds.\cite{Dean2010_BN_graphene,Hattori2015_hBN_breakdown,Rhodes2019_Disorder_vdW} It's Van der Waals nature enables conformal integration with the hydrogen-terminated diamond while minimizing the formation of interface defects. hBN-diamond heterostructures have demonstrated improved electrostatic gating, enhanced current modulation, and stable high-temperature operation, making them relevant for power and quantum electronic applications.\cite{Moon2023_hBN_Photonics,Kim2015_hBN_multilayer,Sasama2018_hBN_HdiamondFET} However, charge transport at the buried hBN/H-terminated diamond interface remains poorly understood, in part because the conduction channel lies beneath an insulating layer that is inaccessible to conventional electrical and scanning probe techniques.\\
Recent advances in nitrogen-vacancy (NV) center magnetometry provide a unique pathway to investigate such buried interfaces. NV ensembles imaged onto a camera enable two-dimensional magnetic field mapping with sub-micrometer spatial resolution and nanoTesla sensitivity over millimeter-scale fields of view.\cite{Rondin2014,Schirhagl2014,LeSage2013,Glenn2015} This technique has been used to image current flow in systems including metallic microstructures,\cite{Schloss2018,Kehayias2019} 2D materials such as graphene,\cite{Tetienne2017_QuantumImagingGraphene} and microcoils for biomedical or microscale electromagnetic device applications.\cite{Simpson2016} Importantly, NV imaging is non-invasive, compatible with high currents and high voltages, and operates at ambient conditions— essential for studying electronic transport in functional devices. Despite these advances, direct magnetic imaging of current flow inside operating diamond FETs has not been reported. The ability to visualize current pathways beneath an hBN dielectric would provide critical insights into the role of interface states, charge screening, non-uniform gating, and current localization in H-terminated diamond devices. Furthermore, diamond FETs integrated on NV-containing substrates present unique opportunities for in situ magnetic imaging without requiring external sensors or invasive probes.\\
In this work, we have fabricated diamond-FETs with hBN gate dielectric to perform spatial mapping of current flow through gated and ungated regions underneath hBN of the device using wide-field magnetic imaging. The devices were fabricated directly on diamond substrate with NV-layer to ensure sub-micron proximity between the 2DHG channel and the NV sensing plane. By capturing gate-voltage-dependent magnetic field maps while simultaneously recording I–V characteristics, we directly correlate spatial current distributions with device transfer characteristics. During electrical characterization, we observe a pronounced enhancement of the channel current under green (532 nm) laser illumination used for continuous-wave optically detected magnetic resonance (CW-ODMR), accompanied by a shift in threshold voltage. This photo-induced modulation reflects optically driven changes in the surface charge environment, with NV photoionization in the buried sensing layer providing a direct source of excess holes that accumulate at the hydrogen-terminated surface and enhance the two-dimensional hole gas density.
Together, these measurements enable a direct correlation between spatially resolved current distributions and electrical transport characteristics, providing new insights into buried interface transport, non-uniform gating effects, and the coupling between optical excitation and surface charge dynamics in diamond FETs. Our demonstration establishes wide-field NV magnetometry as a powerful tool for studying charge transport in diamond electronic devices and, more broadly, for probing buried interfaces in wide-bandgap and Van der Waals heterostructures.

\section{Results and discussion}

\subsection{Experimental setup and device fabrication}
\begin{figure}[]
	\centering
	\includegraphics[width=\textwidth]{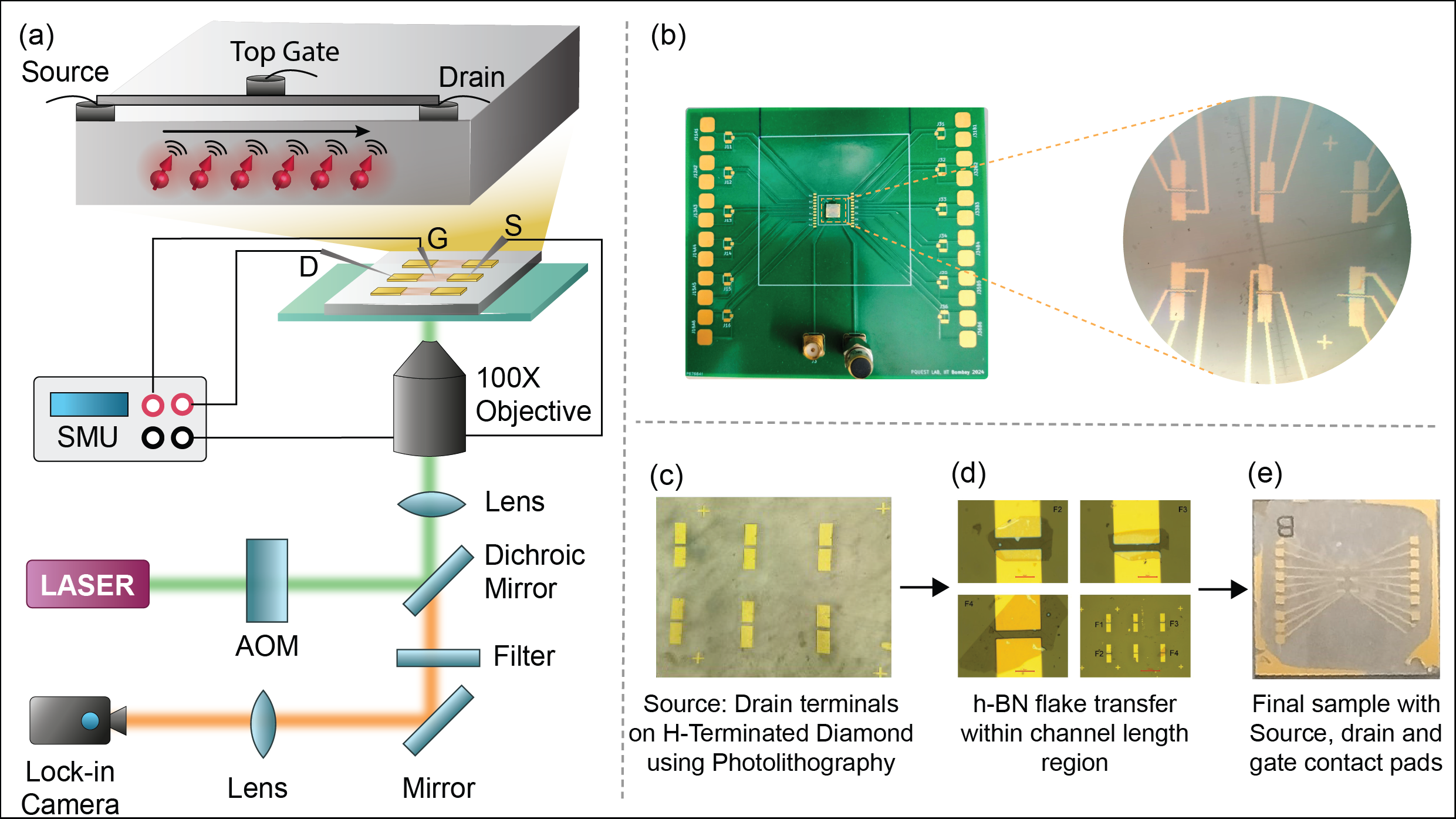} 
	\caption{\textbf{Experimental Setup and Device} (a) Schematic of the wide-field quantum diamond microscope used for NV-based magnetic imaging. (Not drawn to scale) (b) PCB integrated with microwave loop antenna-mounted diamond device (inset: optical image of the fabricated devices). (c) Image of patterned source and drain electrodes defining the hydrogen-terminated channel. (d) Image of devices with hBN flake onto the channel region as the gate dielectric using dry transfer method. (e) Final hBN-gated diamond FET device with source, drain, and gate contact pads.}
	\label{fig1} % give each figure a logical label name
\end{figure}
Wide-field NV magnetometry provides a noninvasive means to visualize current flow in electronic devices with sub-micrometer spatial resolution under ambient conditions. Figure \ref{fig1}(a) illustrates the wide-field quantum diamond microscope (QDM) used in this study. The system operates using CW-ODMR with lock-in detection, enabling sensitive mapping of the static magnetic fields generated by electrical currents. A 532-nm laser beam is expanded to uniformly illuminate the NV layer of the diamond, which is mounted on a custom-designed PCB integrating a microwave loop antenna (see Figure \ref{fig1}(b)). The resulting NV photoluminescence (PL) is collected and imaged onto a scientific CMOS lock-in camera, from which pixel-wise ODMR spectra are extracted via frequency-resolved lock-in demodulation\cite{TemporalSciRep2022}. This measurement scheme suppresses technical noise and enables accurate determination of resonance shifts that encode the local magnetic field. The magnetic field component projected along a selected NV axis is obtained from the differential shift in resonance frequency between current-on and current-off states, referenced to an applied static bias field. The magnetic field of view is $ \sim 100 \mu m$ while maintaining a spatial resolution of $\sim 1 \,\mu\text{m}$, allowing direct visualization of the mesoscale current flow throughout the device.

To enable direct magnetic imaging of the conduction channel, all devices were fabricated on a diamond substrate containing an NV sensing layer located $\sim 1 \,\mu\text{m}$ beneath the surface. The diamond surface was hydrogen terminated to form a conductive two-dimensional hole gas (2DHG), after which source and drain electrodes were patterned using photolithography (Microwriter ML3) followed by wet etching, defining a $ 10 \mu m$ long H-terminated channel between the contacts(Figure \ref{fig1}(c)). Ti/Au metal pads were deposited via an Orion sputtering system to form low-resistance source/drain contacts. Exfoliated hexagonal boron nitride (hBN) flakes were then transferred onto selected channels using a dry-transfer process to serve as high-quality gate dielectric(Figure \ref{fig1}(d)). Top-gate, drain, and source electrodes were subsequently patterned and electrically connected to the global pads, yielding the final device. The chip was then mounted on the custom PCB (Figure 1b,e). Fabricating the devices directly on the NV side minimizes the standoff between the current-carrying 2DHG and the sensing layer, which is essential for achieving high magnetic contrast. In total, six devices were fabricated on the diamond substrate: four that incorporate hBN flakes to form fully gated FETs and two that are without hBN. The complete fabrication workflow is provided in the Supporting Information.

\subsection{Electrical characterization and laser-induced modulation}
\begin{figure}[]
	\centering
	\includegraphics[width=\textwidth]{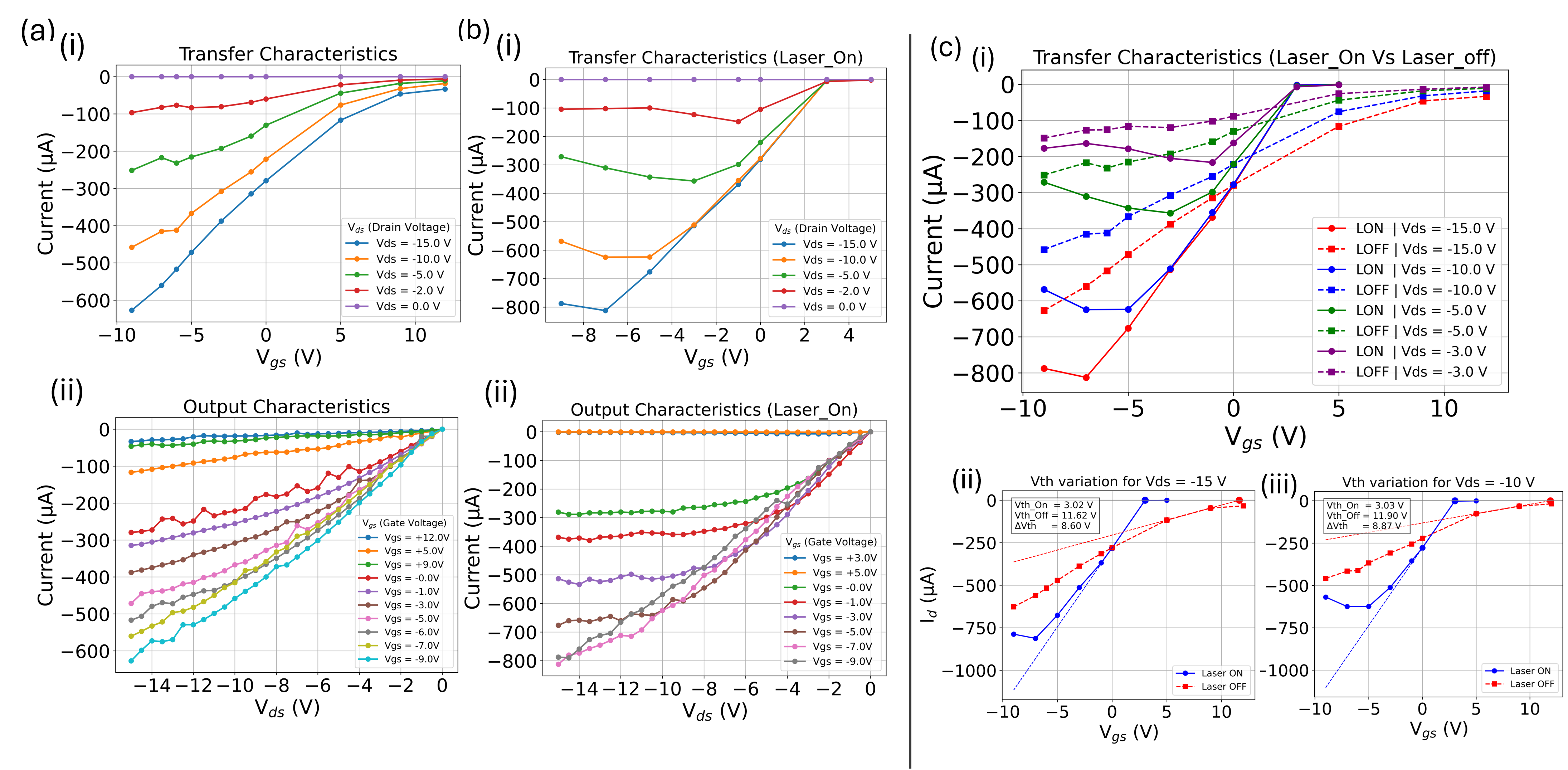} 
	\caption{\textbf{Electrical characterization and laser-induced modulation hBN–diamond FET.} (a) Transfer characteristics measured without laser illumination. (b) Corresponding characteristics under 532 nm laser excitation, showing enhanced drain current. (c) Comparison of laser-on and laser-off transfer curves and extracted threshold voltage shift, indicating photo-induced enhancement of the channel conductivity}

	\label{fig2} % give each figure a logical label name
\end{figure}

Electrical characterization of the hBN–diamond FETs was performed on the wide-field NV magnetometry platform using a custom-built probe setup that enabled simultaneous electrical biasing and NV optical measurements. Tungsten micromanipulator probes were used as contacts for source, drain, and gate pads, enabling controlled application of gate-source voltage ($V_{gs}$) and source-drain voltage ($V_{ds}$) while maintaining stable ODMR excitation and readout conditions.\\
Figures \ref{fig2}a(i) and \ref{fig2}a(ii) show the transfer ($I_{d}$–$V_{gs}$) and output ($I_{d}$–$V_{ds}$) characteristics measured in the absence of laser illumination. The devices exhibit the expected transport behavior of a H-terminated diamond channel, with the drain current increasing monotonically as $V_{gs}$ become more negative and saturating at larger negative $V_{ds}$. The positive threshold voltage indicates depletion-mode operation, consistent with the presence of a surface-transfer-doped 2DHG that forms a conductive channel at zero gate bias. These measurements establish the baseline electrical response of the devices and confirm stable FET operation.\\
Under continuous 532-nm laser illumination—required for NV spin readout—the electrical characteristics undergo a pronounced modification, as shown in Figures \ref{fig2}b(i) and (ii). The drain current increases across the full gate-voltage range, with the peak current rising from approximately $600\,\mu\mathrm{A}$ (laser off) to nearly $900\,\mu\mathrm{A}$ (laser on). At the most negative gate bias ($V_{gs}\approx -8V$) a slight reduction in current is observed in the high-current regime, which may indicate the onset of self-heating effects arising from increased power dissipation and a corresponding reduction in carrier mobility. A direct comparison of the transfer characteristics without and with illumination (Figure \ref{fig2}(c)) further reveals a substantial shift of the apparent threshold voltage toward smaller positive values, with $\Delta V_{th} \approx 8.60 V$ at $V_{ds}= -15V$ and and a comparable shift for $V_{ds}= -10V$.\\
This laser-induced modulation reflects optically driven changes in the electrostatic environment of the channel rather than a simple shift of the transfer curve. Optical excitation of the buried NV layer can introduce additional mobile charge through NV charge-state cycling, while illumination may also modify surface and dielectric charge distributions. Together, these effects reduce the gate bias required to electrostatically deplete the channel, leading to the observed shift in threshold voltage. At the same time, the enhanced drain current under illumination indicates an overall increase in channel conductivity. A quantitative analysis of photo-induced carrier generation mechanisms in NV-layer diamond, and their consistency with the observed electrical response, is presented in the Supporting Information.

\subsection{Wide-field magnetic imaging and reconstruction of current paths}
\begin{figure}[]
	\centering
	\includegraphics[width=\textwidth]{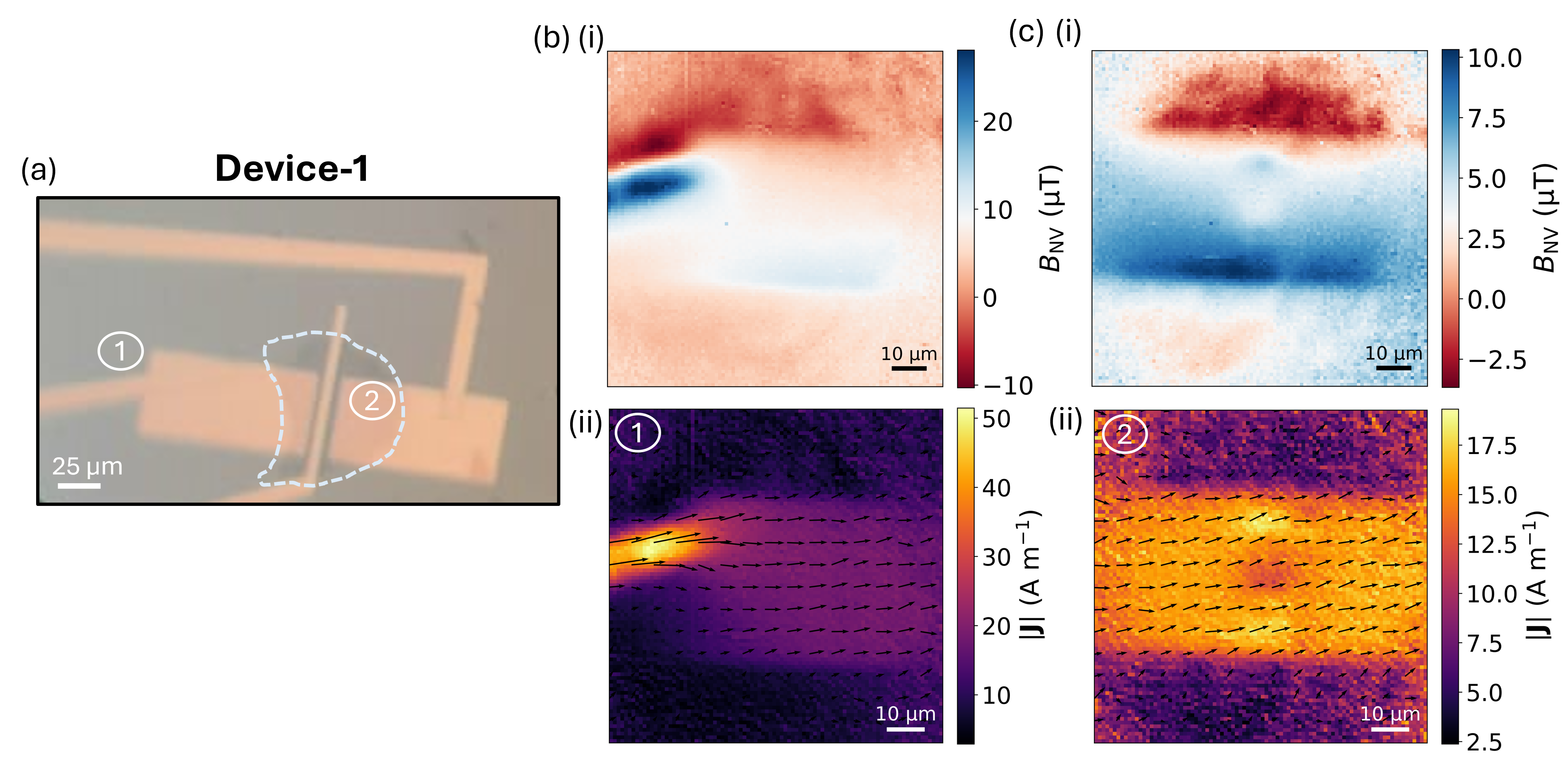} 
	\caption{\textbf{Wide-field magnetic imaging and current-density reconstruction in hBN–diamond FETs} (a) Optical micrograph of Device~1 highlighting the source--drain contacts (\textcircled{1}) and the hBN-gated channel (\textcircled{2}); the dashed outline indicates the boundary of the hBN flake.(b,c) Magnetic field maps $B_{NV}$ and corresponding reconstructed current density $|\mathbf{J}|$ at the contact interface and gated channel, respectively, measured at $V_{ds}= -12V$,$V_{gs}= -5V$}
	\label{fig3} % give each figure a logical label name
\end{figure}

To visualize the spatial distribution of current flow in operating hBN–diamond FETs, we employed wide-field quantum diamond microscopy to image the magnetic fields generated by the channel current and reconstruct the corresponding two-dimensional current density. Figure \ref{fig3} shows the magnetic imaging and current reconstruction results for Device-1, which incorporates a uniform hBN gate dielectric; details of the magnetic-field–to–current-density reconstruction procedure and the associated regularization scheme are provided in the Supporting Information.
Figure \ref{fig3}(a) shows an optical image of Device-1, highlighting the two regions investigated: (i) the source–drain contact interface and (ii) the hBN-gated channel region. Magnetic field maps were acquired at a fixed operating point of $V_{ds}= -12V$,$V_{gs}= -5V$ under which the device exhibits stable conduction.\\
In the contact region[Fig. \ref{fig3}(b-i)], the magnetic image reveals strong, spatially localized magnetic field lobes characteristic of current crowding, where carriers are injected from the metal contact into the hydrogen-terminated channel. This behavior is consistent with the expected spreading resistance at the contact edge, where the injected current rapidly redistributes into the two-dimensional hole gas (2DHG). The reconstructed current density map [Fig. \ref{fig3}(b-ii)] confirms this interpretation, revealing a concentrated injection hotspot that gradually broadens as the current redistributes laterally into the two-dimensional hole gas (2DHG).
In the hBN-gated channel region [Fig. \ref{fig3}(c-i)], the magnetic field map displays a smooth, spatially continuous profile extending across the channel width. The corresponding current density reconstruction [Fig. 3(c-ii)] reveals a well-distributed and nearly uniform current flow beneath the hBN dielectric, with current streamlines aligned along the channel direction. This uniformity indicates effective electrostatic control and homogeneous gating provided by the high-quality hBN dielectric, consistent with the stable transfer characteristics observed electrically. Together, these spatially resolved measurements directly relate the local magnetic response to the internal current pathways of the device, revealing that Device-1 supports highly uniform gating and transport across both contact and channel regions.\\
For comparison, Device-2, which incorporates a non-uniform hBN flake, exhibits markedly different behavior, as shown in Fig.\ref{fig7} of the Supporting Information. In that case, both the magnetic field and reconstructed current density displayed pronounced spatial modulation, with localized regions of enhanced and reduced field amplitude. These variations directly trace back to the non-uniformity of the transferred hBN flake. Thicker or partially delaminated regions of hBN weaken the local gate coupling, allowing higher current densities to pass through those regions, while thinner or better-adhered patches suppress current more effectively. The resulting pattern, therefore, indicates the local variability in gate control of the channel due to defects in the dielectric. These results further highlight the critical role of dielectric uniformity in governing current transport in hBN–diamond FETs.

Overall, wide-field NV magnetometry enables direct, noninvasive visualization of current flow beneath buried dielectrics, providing insights into contact injection, gate-induced current redistribution, and dielectric-induced transport inhomogeneity that remain inaccessible to standard electrical measurements alone.

\subsection{I-V mapping with hBN gate dielectric interface magnetic imaging}
\begin{figure}[]
	\centering
	\includegraphics[width=\textwidth]{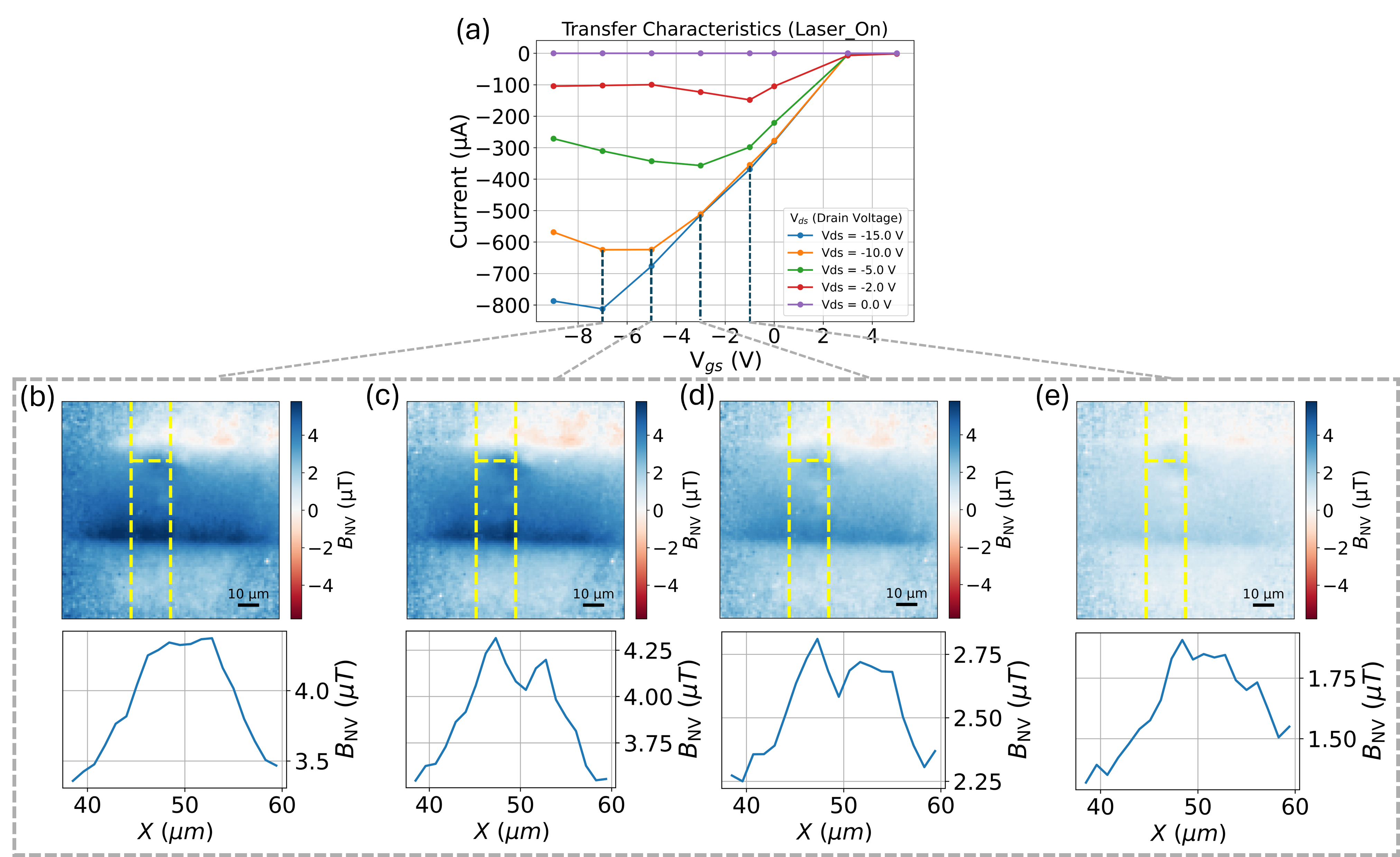} 
	\caption{\textbf{Gate-dependent magnetic field imaging correlation with electrical transport} (a) Transfer characteristics of the diamond FET at $V_{ds}= -10V$ showing the increase in drain current as $V_{gs}$ becomes more negative.(b-e) Wide-field NV magnetic field maps ($B_{NV}$ )acquired at the corresponding gate voltages. The line-cut plots shown below each image are horizontal profiles, revealing how the spatial current distribution evolves with electrostatic gating.}
	\label{fig4} % give each figure a logical label name
\end{figure}

To directly correlate the electrical behaviour of the diamond FET with its internal current pathways, wide-field NV magnetic imaging was performed while sweeping the gate voltage at fixed drain bias of $V_{ds}= -10V$. Figure \ref{fig4}(a) shows the corresponding transfer characteristics, where the magnitude of the drain current systematically increases as $V_{gs}$ becomes more negative, consistent with p-type 2DHG conduction.  Each selected operating point is linked to a magnetic field map in Figures \ref{fig4}(b–e), establishing a direct correspondence between the I–V response and the spatial magnetic images and enabling the gating dynamics to be visualized inside the channel.\\
The magnetic images reveal that the overall magnitude of $B_{NV}$ decreases as the channel current reduces with gate sweep. Beyond this global trend, the spatial evolution of the magnetic signal reveals nonuniform current suppression across the channel. In particular, regions near the edges of the hBN flake exhibit earlier and stronger attenuation of the magnetic field compared to the central channel region, reflecting spatial variations in local gate efficiency caused by nonuniform dielectric thickness or imperfect hBN–diamond interface coupling. Together, these measurements demonstrate that concurrent electrical characterization and wide-field NV magnetometry provide a powerful framework for visualizing how electrostatic gating reshapes current flow inside diamond FETs, revealing spatial inhomogeneities in dielectric coupling and transport that are inaccessible to conventional electrical measurements.

\subsection{Conclusion and Outlook}
In this work, we have demonstrated, for the first time, wide-field magnetic imaging of current flow in operating hBN-gated hydrogen-terminated diamond FETs using an ensemble NV center quantum diamond microscope. By fabricating the devices directly on an NV-layer diamond substrate, we achieved sub-micrometer proximity between the current-carrying two-dimensional hole gas and the sensing layer, enabling noninvasive, spatially resolved visualization of current injection, redistribution, and gating effects under realistic bias conditions.

Correlating electrical transport measurements with concurrent magnetic field imaging allowed us to directly connect macroscopic device characteristics to microscopic current pathways. Magnetic maps revealed clear signatures of current crowding at the source–drain contacts and gate field-dependent redistribution of current beneath the hBN dielectric, with pronounced spatial inhomogeneities arising from non-uniform dielectric coverage. By reconstructing the two-dimensional current density from multi-axis NV measurements, we obtained quantitative current maps that provide direct insight into how gate dielectrics modulate carrier transport in buried diamond interfaces—providing information that remains inaccessible through conventional electrical measurements alone.
A key finding of this study is the pronounced laser-induced enhancement of channel conductivity observed during NV magnetometry measurements. Through a quantitative model supported by independent experimental data, we show that this effect originates from photo-induced electron–hole pair generation in the buried NV layer, followed by hole accumulation in the hydrogen-terminated surface channel. The resulting increase in 2DHG density quantitatively accounts for the observed threshold-voltage shift and drain-current enhancement. This analysis highlights the dual role of the NV layer as both a magnetic sensor and an optically addressable carrier reservoir, an effect that must be carefully considered when combining optical magnetometry with active electronic devices.\\
In conclusion, we demonstrate an operando, noninvasive current-imaging methodology based on wide-field NV magnetometry that enables direct, spatially resolved probing of buried charge transport in diamond and Van der Waals' heterostructure devices. This approach provides unique access to transport phenomena that are inaccessible to conventional electrical measurements, including contact resistance, dielectric nonuniformity, and gate-induced inhomogeneities. By performing gate-dependent magnetic imaging, we show the ability to capture dynamic and nonlinear transport processes, such as channel pinch-off, percolation, and localized breakdown, in real space. We further demonstrate that engineering the NV layer depth and optical excitation conditions enables optical modulation of carrier density, introducing a new mechanism for coupling spin defects to surface transport and opening pathways toward hybrid optoelectronic device concepts. Collectively, our results establish wide-field NV magnetometry as a general, quantitative, and a platform for noninvasive current imaging across a broad range of emerging electronic materials and device architectures, with significant implications for the design, diagnostics, and optimization of next-generation nanoelectronic systems.

\begin{acknowledgement}

We thank Prof Padmanabh Rai from CEBS, Mumbai University for enabling H-termination on the diamond for making the devices. K.S. acknowledges funding from DST National Quantum Mission, TCS Research and  Asian Office of Aerospace Research and Development (AOARD) grant number FA2386-23-1-4012. A.B. and K.S. acknowledge support from IITB Nanofabrication facility. A.B. acknowledges funding from I-Hub Quantum Technology Foundation's Chanakya Doctoral Fellowship. A.S. acknowledges funding from MoE-STARS STARS-2/2023-0265. 

\end{acknowledgement}

\begin{suppinfo}
\section{Experimental setup}
Wide-field magnetic imaging of the hBN–diamond MOSFETs was performed using NV-based quantum diamond microscope describe in Figure 1(a), with modifications to enable simultaneous electrical biasing and optical readout. The devices were fabricated directly on the side of the NV-layer of an electronic-grade single-crystal diamond containing a deep layer of $\sim 1 \,\mu\text{m}$ NV centers (NV$^-$) concentration ($\sim 1-2 \, ppm$). The diamond chip was mounted onto a custom PCB incorporating microwave loop antenna to deliver resonant microwave excitation near 2.87 GHz.\\
A 532-nm continuous-wave laser $(\sim 0.4 W)$ was expanded and directed through a $100 \times$ air objective (NA = 0.9) to uniformly illuminate the NV layer beneath the device. The resulting NV photoluminescence was collected through the same objective, spectrally filtered (Semrock NF03-532E-25), and imaged onto a wide-field lock-in CMOS camera (Heliotis Helicam C3). Frequency-modulated microwave excitation was synchronized with the camera using TTL pulse sequences generated by a SpinCore PulseBlaster ESR-PRO (500 MHz), enabling pixel-wise extraction of ODMR contrast and local magnetic field values. The optical system provided 36× magnification and a field of view of approximately $\ 100 \,\mu\text{m}$ × $\ 100 \,\mu\text{m}$, sufficient to capture both contact pads and gated channel regions of the MOSFET.\\
To electrically operate the devices during magnetic imaging, a custom-built microprobe station was integrated into the optical platform. Three tungsten micromanipulator probes were positioned on the source, drain, and gate pads of the MOSFET. A Keithley SourceMeter Unit (SMU 2401) supplied the drain-source bias $V_{ds}$ and simultaneously measured the drain current $I_{ds}$ while a Keithley 2231A power supply provided the gate bias $V_{gs}$. This configuration enabled stable and repeatable electrical biasing during continuous laser illumination and microwave excitation, ensuring that the NV measurements captured the true current distribution corresponding to each electrical operating point.A static bias magnetic field was applied using Sm–Co permanent magnets to lift the degeneracy of the NV spin transitions and improve ODMR contrast. All measurements were performed under ambient laboratory conditions.

\section{Device Fabrication}
The hBN–diamond FETs were fabricated on an electronic-grade single-crystal (100) diamond substrate containing a near-surface NV layer located approximately $\ 1 \,\mu\text{m}$ below the surface. Before fabrication, the diamond substrate was cleaned to remove organic and metallic contaminants using a standard tri-acid treatment $(\ce{H2SO4}:\ce{HNO3}:\ce{HClO4} = 1:1:1)$ followed by sequential rinsing in acetone, isopropanol and deionized water.\\
\begin{figure}[]
	\centering
	\includegraphics[width=\textwidth]{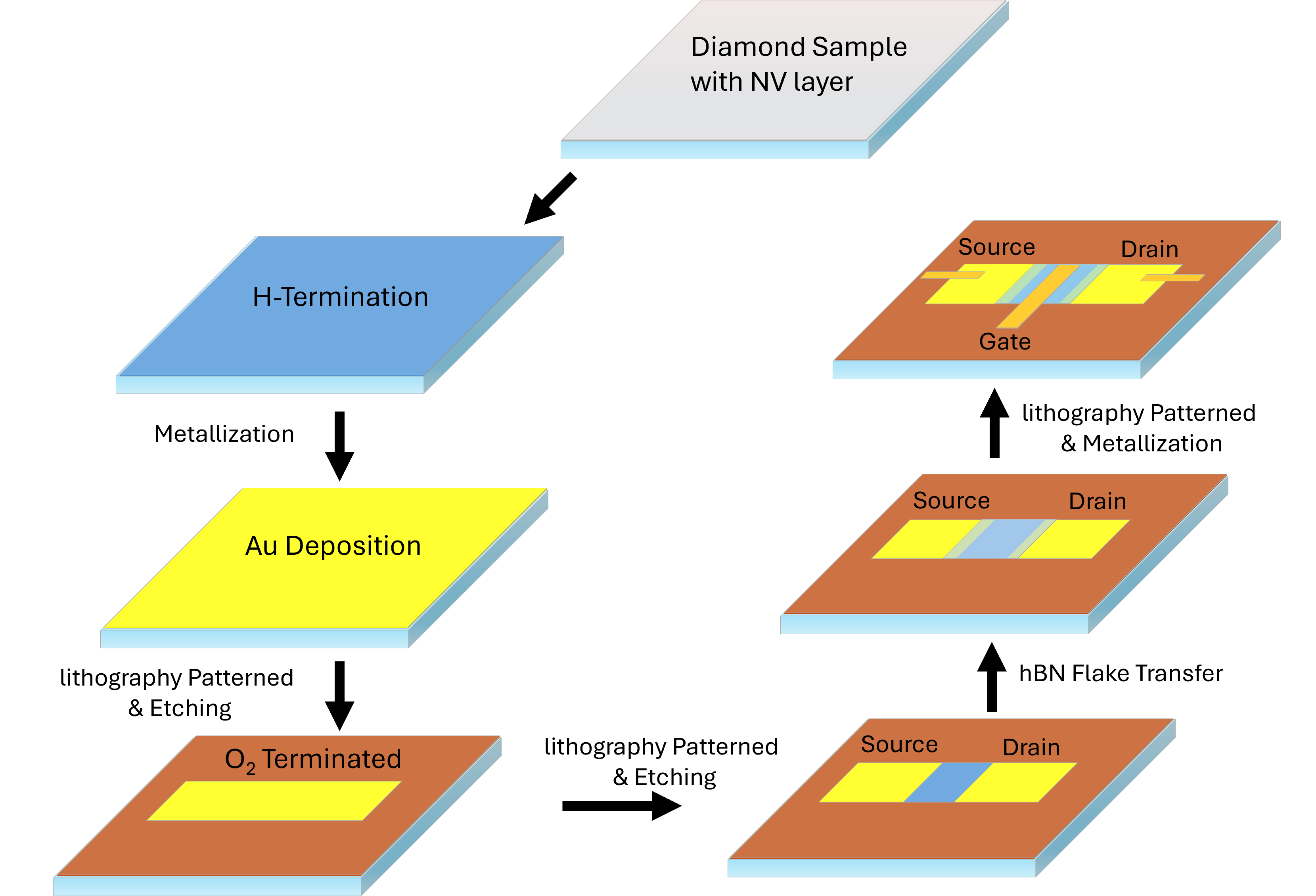} 
	\caption{\textbf{Fabrication process flow for hBN-gated diamond FETs}}
	\label{fig5} % give each figure a logical label name
\end{figure}
\newpage
Hydrogen termination of the diamond surface was performed in a microwave plasma chemical vapor deposition (MPCVD) system. The sample was exposed to hydrogen plasma for 30–40 minutes at a microwave power of 2 to 3 kW, a chamber pressure of 700 Torr and a substrate temperature of $700-750^\circ \text{C}$. During cooling, a hydrogen flow of 100–500 sccm was maintained to stabilize the surface termination. Upon exposure to ambient air for 12–24 hours, a two-dimensional hole gas (2DHG) spontaneously formed through the surface-transfer doping mechanism, in which atmospheric adsorbates remove electrons from the diamond valence band, generating mobile holes confined to the topmost atomic layers.\\
To protect the hydrogen-terminated surface during lithographic processing, a blanket Au layer (100 nm) was deposited using electron-beam evaporation. Device isolation was then achieved using standard photolithography, followed by selective Au etching in a potassium iodide/iodine solution. In regions requiring complete carrier depletion, the exposed diamond was treated with oxygen plasma $(\ce{O2} flow: 90 sccm, RF power: 40–60 W, duration: 2 min)$, which replaces surface hydrogen with oxygen-containing groups and thus fully removes 2DHG to electrically isolate device regions.\\
The source and drain regions were defined in a subsequent lithography step, followed by selective Au wet etching to expose the H-terminated channel, resulting in a channel length $\sim 10 \,\mu\text{m}$. The exfoliated hexagonal boron nitride (hBN) flakes ($\sim20–30$  nm thick) were then transferred to the channel using a dry-transfer method, providing an atomically flat and chemically inert dielectric that avoids the high-temperature or reactive processes associated with conventional oxide deposition\cite{CastellanosGomez2014DeterministicTransfer}.
A final lithography and lift-off step defined the gate electrode on top of the hBN dielectric. A Ti/Au metal stack (30 nm / 100 nm) was deposited to form the gate and extended into large contact pads suitable for probing and PCB interfacing during NV magnetic imaging.

\section{Current Density Reconstruction from Wide-Field NV Magnetic Imaging}

To obtain spatially resolved maps of current flow in the hBN--diamond FETs, we reconstructed
the two-dimensional current density $\mathbf{J}(x,y)$ from magnetic field maps measured by
the NV ensemble. The reconstruction procedure involved four stages:
(i) multi-axis NV magnetic imaging,
(ii) determination of NV orientations in the laboratory frame,
(iii) recovery of the full vector magnetic field, and
(iv) application of a joint-optimization-based current reconstruction algorithm.

Magnetic field maps were acquired along three accessible NV crystallographic orientations
(NV$_1$, NV$_2$, NV$_3$). For each orientation, differential ODMR measurements were performed
under current-on and current-off conditions, yielding the projected magnetic field
\begin{equation}
B_{\mathrm{NV}_i}(x,y) = \hat{\mathbf{n}}_i \cdot \mathbf{B}(x,y),
\end{equation}
where $\hat{\mathbf{n}}_i$ is the unit vector associated with the corresponding NV axis. Because the measured projections cannot be directly assigned to specific crystallographic axes, the correct NV orientation must be determined before reconstructing the full vector magnetic field \( (B_x, B_y, B_z) \).\\
To identify the NV axes, we computed the magnetic field expected for our device geometry using COMSOL Multiphysics. A simplified MOSFET model—with the same current magnitude but without the gate dielectric—was simulated to obtain \( (B_x', B_y', B_z') \).  
These simulated components were projected onto all four possible \( \langle 111 \rangle \) NV orientations:
\[
B_{\mathrm{NV}_i}' = 
\hat{n}_i \cdot 
\begin{pmatrix}
B_x' \\ B_y' \\ B_z'
\end{pmatrix}.
\]
Figure.6 compares these simulated projections to the experimentally measured $B_{\mathrm{NV}_i}$ maps. The one-to-one correspondence between spatial features in simulated and measured data uniquely identifies NV\(_1\), NV\(_2\) and NV\(_3\)in the laboratory frame. With the NV orientations known, the system.

\[
\begin{pmatrix}
B_{\mathrm{NV}_1} \\
B_{\mathrm{NV}_2} \\
B_{\mathrm{NV}_3}
\end{pmatrix}
=
\begin{pmatrix}
\hat{n}_{1x} & \hat{n}_{1y} & \hat{n}_{1z} \\
\hat{n}_{2x} & \hat{n}_{2y} & \hat{n}_{2z} \\
\hat{n}_{3x} & \hat{n}_{3y} & \hat{n}_{3z}
\end{pmatrix}
\begin{pmatrix}
B_x \\ B_y \\ B_z
\end{pmatrix}.\]
was solved at each pixel to recover the full vector magnetic field \[ B_x(x,y),\, B_y(x,y),\, B_z(x,y).\]
These Cartesian fields served as input to the current reconstruction. The inverse problem is intrinsically ill-posed—particularly when the NV–sample standoff distance is comparable to or larger than the spatial sampling scale (i.e., large standoff-to-pixel-size ratio), leading to strong suppression of high-spatial-frequency components of the magnetic field\cite{roth1989using}and therefore requires a regularized reconstruction framework, as introduced in our earlier work\cite{Anand2025JointEstimationQDM}. The forward model relating the current density to the magnetic field is given by:
\[
B_x = G_3 * J_y, \qquad
B_y = -\, G_3 * J_x, \qquad
B_z = G_1 * J_x - G_2 * J_y,
\]
where \( G_1, G_2, G_3 \) are Green's function kernels derived from the Biot--Savart law. 
The current density is obtained by solving the joint optimization problem:
\[
\arg\min_{J_x, J_y}
\left[
\begin{aligned}
&\frac{w_1}{2}\, \| B_x - G_3 * J_y \|_2^2 \\
&+\frac{w_2}{2}\, \| B_y + G_3 * J_x \|_2^2 \\
&+\frac{w_3}{2}\, \| B_z - (G_1 * J_x - G_2 * J_y) \|_2^2 \\
&+\lambda_1 \| \nabla \cdot \mathbf{J} \|_2^2 \\
&+\lambda_2\, R(J_x, J_y)
\end{aligned}
\right].
\]
where the first three terms enforce agreement with the measured magnetic fields, 
the divergence penalty ensures physical current continuity, 
and \( R \) is a BM3D-based denoising prior that suppresses noise-induced artifacts. 
The optimization is solved using an alternating direction method of multipliers(ADMM) based iterative scheme, yielding stable and high-contrast current-density maps even at low signal-to-noise ratios.

\begin{figure}[]
	\centering
	\includegraphics[width=\textwidth]{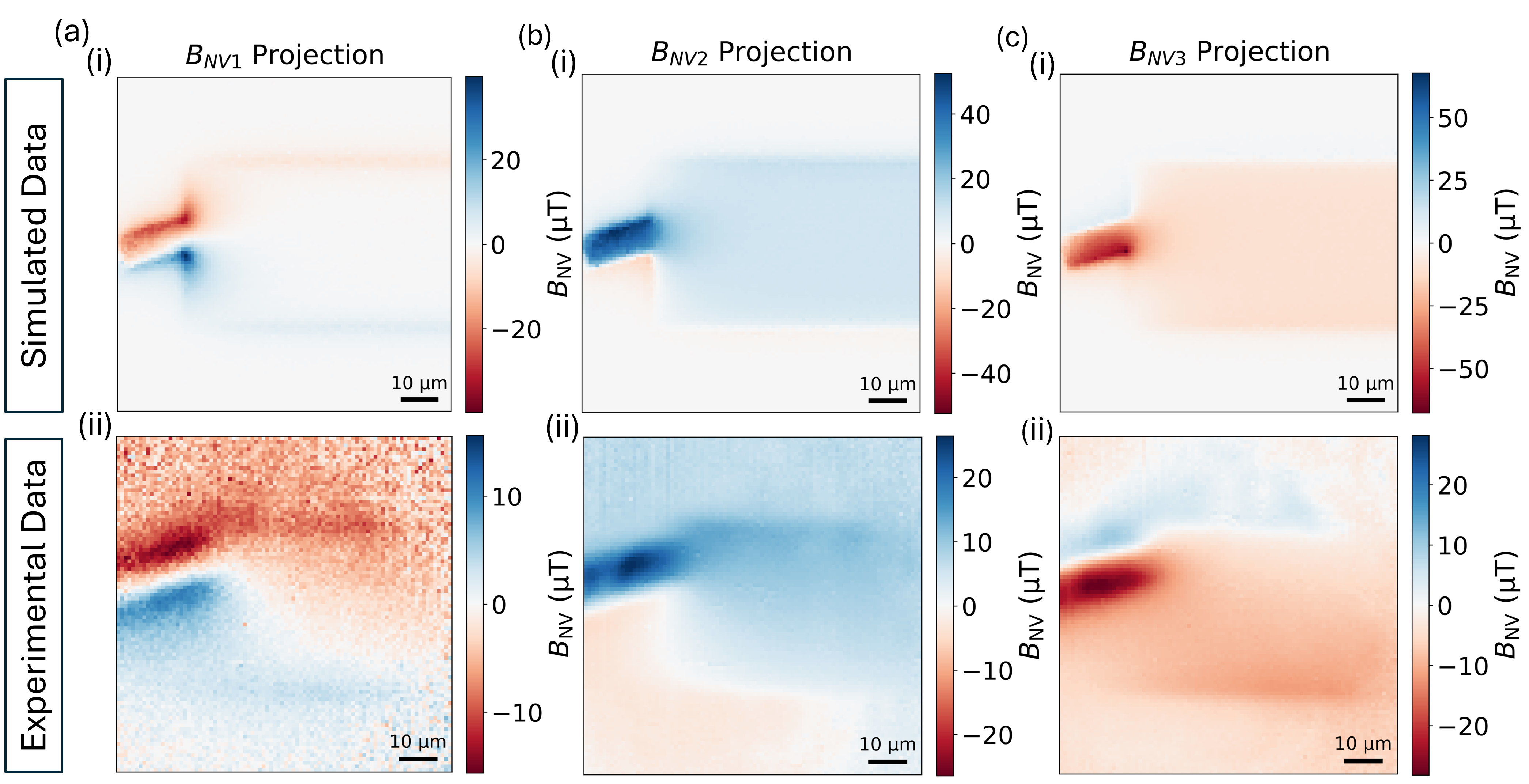} 
	\caption{\textbf{NV-axis identification for vector magnetic field reconstruction} Simulated (top row) and experimentally measured (bottom row) magnetic projections for three NV orientations. Matching the spatial patterns between simulation and experiment uniquely determines the $NV_1$, $NV_2$, and $NV_3$ axes in the laboratory frame, enabling reconstruction of the full vector magnetic field used for current-density mapping.}
	\label{fig6} % give each figure a logical label name
\end{figure}
\newpage
\begin{figure}[]
	\centering
	\includegraphics[width=\textwidth]{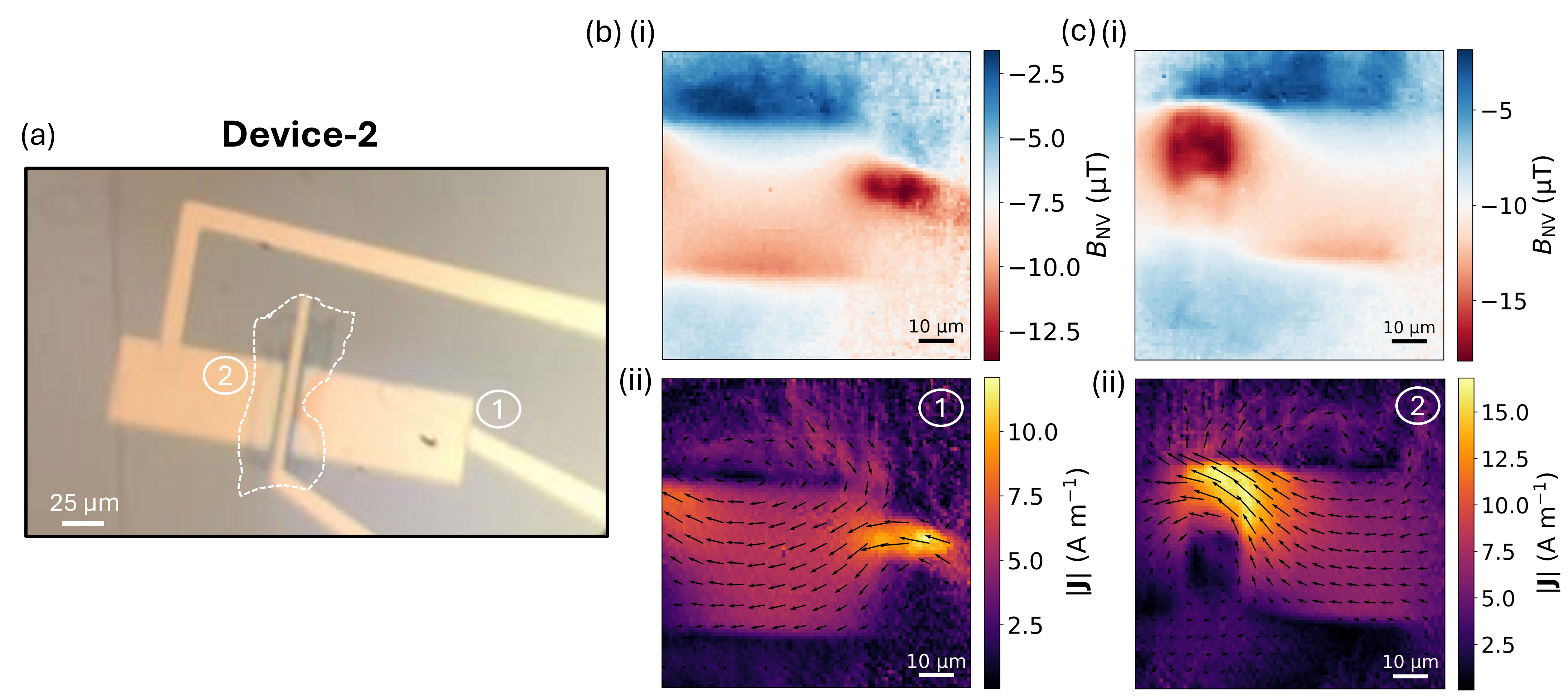} 
	\caption{\textbf{Wide-field magnetic imaging and current-density reconstruction for Device-2 with a non-uniform hBN gate dielectric}(a) Optical micrograph of Device~1 highlighting the source--drain contacts (\textcircled{1}) and the hBN-gated channel (\textcircled{2}); the dashed outline indicates the boundary of the hBN flake.
(b,c) Magnetic field maps $B_{NV}$ and corresponding reconstructed current density $|\mathbf{J}|$ acquired at ($V_{ds}= -19V$,$V_{gs}= -5V$) showing spatially nonuniform current flow arising from variations in local gate coupling.}
	\label{fig7} % give each figure a logical label name
\end{figure}

\section{Photo-Induced Conductivity Enhancement in hBN--\\Diamond FETs}
The hBN–diamond FETs exhibit a pronounced increase in drain current and a shift in the apparent threshold voltage under 532 nm laser illumination used for NV magnetometry. In this section, we develop a quantitative model to evaluate whether photo-induced carrier generation in the buried NV layer can provide a sufficient contribution to the observed electrical modulation. The analysis demonstrates that NV-mediated carrier generation is quantitatively consistent with the experimentally observed changes.

\subsection*{NV-Mediated Carrier Generation Mechanism}

Optical excitation of NV centers drives a charge-state cycle involving photoionization and recharging processes:\\

\begin{equation}
\mathrm{NV^-} + 2h\nu \rightarrow \mathrm{NV^0} + e^- ,
% \label{eq:1}
% \tag{1}
\end{equation}
\begin{equation}
\mathrm{NV^0} + h\nu \rightarrow \mathrm{NV^-} + h^+ ,
% \label{eq:S2}
% \tag{2}
\end{equation}
Each complete NV charge-state cycle results in the generation of one electron–hole pair. The photo-generated electrons are trapped or recombine in the bulk, while holes can diffuse toward the hydrogen-terminated diamond surface and influence the electrostatics of the two-dimensional hole gas (2DHG)\cite{Aslam_NV_Ionization,Le2025_FEDMR_NV}.

\subsection*{NV population dynamics under illumination}

Let $n^-$ and $n^0$ denote the densities of $\mathrm{NV^-}$ and $\mathrm{NV^0}$ centers, respectively. The total NV density is
\begin{equation}
    n_{\mathrm{NV}} = n^- + n^0 .
    %\tag{3}
\end{equation}

Under optical illumination with photon flux $\Phi$, the charge-state dynamics defined by the rate equation
\begin{equation}
    \frac{dn^-}{dt}
    = -\Gamma_{\mathrm{ion}} \, n^-
    + \Gamma_{\mathrm{rec}} \, n^0 ,
    %\tag{4}
\end{equation}
where $\Gamma_{\mathrm{ion}}$ and $\Gamma_{\mathrm{rec}}$ are the effective photo-ionization and photo-assisted recombination rates, respectively.

At steady state,
\begin{equation*}
    \frac{dn^-}{dt} = 0 ,
\end{equation*}
the population ratio is
\begin{equation}
    \frac{n^0}{n^-}
    = \frac{\Gamma_{\mathrm{ion}}}{\Gamma_{\mathrm{rec}}} .
    %\tag{5}
\end{equation}

Experimental studies of NV charge-state dynamics under continuous-wave green excitation report that recombination dominates over ionization, such that
\begin{equation*}
    \Gamma_{\mathrm{ion}} \ll \Gamma_{\mathrm{rec}} ,
\end{equation*}
and therefore
\begin{equation}
    n^- \approx n_{\mathrm{NV}} .
    %\tag{6}
\end{equation}

\subsection*{For the given experimental parameters}

\paragraph{Diamond sample}

\begin{itemize}
    \item N concentration: $12~\mathrm{ppm} \approx 2.1 \times 10^{18}~\mathrm{cm^{-3}}$
    \item NV concentration $(n_{\mathrm{NV}})$: $3.8~\mathrm{ppm} \approx6.6 \times 10^{17}~\mathrm{cm^{-3}}$
    \item NV layer thickness: $d = 1~\mu\mathrm{m} = 10^{-4}~\mathrm{cm}$
\end{itemize}

\paragraph{Laser illumination}
\begin{itemize}
    \item Wavelength $(\lambda) = 532~\mathrm{nm} (h\nu = 2.33~\mathrm{eV})$
    \item Optical power at the sample: $P = 0.35$--$0.40~\mathrm{W}$
    \item Illuminated FOV: $d \approx 100~\mu\mathrm{m}$
\end{itemize}

The corresponding photon flux $(\Phi)$ at the NV layer
\begin{equation}
    \Phi = \frac{P}{A h\nu}
    \approx (1\text{--}4)\times 10^{22}~\mathrm{cm^{-2}\,s^{-1}} .
    %\tag{S5.6}
\end{equation}

Based on the experimentally observed quadratic dependence of the $\mathrm{NV^-}$ ionization rate on laser power \cite{Aslam_NV_Ionization},the measured ionization rate per NV center is of order

\begin{equation}
    \Gamma_{\mathrm{ion}} \approx 10^{2}\text{--}10^{3}~\mathrm{s^{-1}} ,
    %\tag{S5.9}
\end{equation}

\subsection{Electron--Hole pair generation rate in the NV layer}
Each NV ionization event produces one electron--hole pair.The volumetric generation rate within the NV layer is therefore
\begin{equation}
    G_{\mathrm{eh}} = \Gamma_{\mathrm{ion}} \, n^-
    \approx \Gamma_{\mathrm{ion}} \, n_{\mathrm{NV}} ,
    %\tag{S5.10}
\end{equation}
which yields
\begin{equation}
    G_{\mathrm{eh}} \approx 10^{20}\text{--}10^{21}~\mathrm{cm^{-3}\,s^{-1}} .
    %\tag{S5.11}
\end{equation}
Photo-generated holes diffuse rapidly to the hydrogen-terminated diamond surface and enter the surface accumulation layer. The steady-state increase in the surface hole sheet density is given by
\begin{equation}
    \Delta p_s = G_{\mathrm{eh}} \, \tau_h \, d ,
    %\tag{S5.12}
\end{equation}
where $\tau_h$ is the effective life time of excess holes, governed by surface recombination and extraction processes in the surface accumulation layer. Using reported surface recombination velocities for diamond and an accumulation layer thickness of a few nanometers\cite{Grivickas2020_CarrierRecombinationDiamond}. Estimated effective hole life time for near-surface diamond,
\begin{equation}
    \tau_h \sim 1\text{--}10~\mu\mathrm{s},
\end{equation}
which yields
\begin{equation}
\boxed{
    \Delta p_s \sim 10^{10}\text{--}10^{11}~\mathrm{cm^{-2}}
    }
    %\tag{S5.13}
\end{equation}

\subsection*{Consistency With Experiment: Threshold Voltage modulation}

For an hBN-gated surface two-dimensional hole gas (2DHG), the threshold voltage is given by
\begin{equation}
    V_{\mathrm{th}} = V_{\mathrm{fb}} - \frac{q\, p_s}{C_{\mathrm{ox}}} ,
\end{equation}
where $p_s$ is the hole sheet density and $C_{\mathrm{ox}}$ is the areal capacitance of the hBN gate dielectric, defined as
\begin{equation}
    C_{\mathrm{ox}} = \frac{\varepsilon_0 \varepsilon_{\mathrm{hBN}}}{t_{\mathrm{hBN}}} .
\end{equation}

Under laser illumination, An effective electrostatic modulation corresponds to
\begin{equation}
    \Delta V_{\mathrm{th}}
    = V_{\mathrm{th,on}} - V_{\mathrm{th,off}}
    = -\frac{q\, \Delta p_s}{C_{\mathrm{ox}}} .
\end{equation}

\begin{equation}
    \Delta p_s
    = \frac{|\Delta V_{\mathrm{th}}|\, C_{\mathrm{ox}}}{q} .
    %\tag{S1}
\end{equation}
Using 
\begin{itemize}
    \item $|\Delta V_{\mathrm{th}}| = 8.6~\mathrm{V}$
    \item $t_{\mathrm{hBN}} = 20~\mathrm{nm} = 2 \times 10^{-8}~\mathrm{m}$
    \item relative permittivity: $\varepsilon_{\mathrm{hBN}} \approx 3.5,$
    \item elementary charge $q = 1.6 \times 10^{-19}~\mathrm{C}$
\end{itemize}

yields an equivalent charge modulation
\begin{equation}
    \boxed{
    \Delta p_s = 8.3 \times 10^{12}~\mathrm{cm^{-2}}
    }.
    %\tag{S5.13}
\end{equation}
This value lies within the upper range of the NV-mediated contribution estimated above, indicating that NV photoionization can provide a substantial contribution to the observed electrostatic modulation. Additional photo-induced charging at the hydrogen-terminated surface and hBN interface may further enhance this effect.

\end{suppinfo}

\bibliography{achemso-demo}

\end{document}